%% file: harness-evolution-design-verification-arxiv.tex
\documentclass{article}

\PassOptionsToPackage{numbers,compress}{natbib}
\usepackage[main,preprint]{neurips_2026}
\usepackage[utf8]{inputenc}
\usepackage[T1]{fontenc}
\usepackage[hidelinks]{hyperref}
\usepackage{url}
\usepackage{booktabs}
\usepackage{tabularx}
\usepackage{graphicx}
\usepackage{float}
\usepackage{amsmath}
\usepackage{amsfonts}
\usepackage{microtype}
\usepackage{xcolor}
\usepackage{pifont}

\newcounter{paperalgorithm}
\newenvironment{paperalgorithm}[1]{%
  \refstepcounter{paperalgorithm}%
  \par\medskip
  \noindent\begin{minipage}{\linewidth}
  \hrule\vspace{0.45em}
  \textbf{Algorithm \thepaperalgorithm: #1}
  \par\vspace{0.20em}
  \small
  \setlength{\parskip}{0.08em}
}{%
  \par
  \vspace{0.2em}\hrule
  \end{minipage}
  \par\medskip
}

\title{Automatic Harness Evolution for Hardware\\
Design Verification: Can LLMs Consolidate\\
Gains Across Discovered Harnesses?}

\author{%
Kidus Seyoum\thanks{Work performed during an internship at NVIDIA.}\\
NVIDIA\\
\texttt{kseyoum@nvidia.com}
\And
Ajay Mittur\thanks{Corresponding author.}\\
NVIDIA\\
\texttt{amittur@nvidia.com}
}

\begin{document}

\maketitle

\begin{abstract}
Agent behavior depends on the harness surrounding a language model, but it remains unclear whether language models can reliably improve such harnesses for hardware-design tasks. We study automatic harness evolution around a fixed subject model on 12 proprietary design-verification root-cause localization tasks. Across five trials per task, automatically evolved harnesses increased completed attempts by 71--76\% and any-hit task coverage by 80--100\%, while total correct attempts improved by only 18--24\%. The strongest Success\(\geq\)2/5 result improved by one task, and later candidates exchanged gains across tasks rather than preserving them. An auxiliary candidate improved on a four-task validation set excluded from search but tied its baseline on a subsequent 12-task replay containing both search and validation tasks, so the selected gain did not persist across the full pool. Across the tested lineage, useful search, evidence, and finalization behaviors appeared in different candidates but did not consistently consolidate into a single harness that dominated across tasks and metrics. In a separate CVDP cross-benchmark case study, an automatically evolved defined-width repair harness produced 35.6\% more functional passes than its 142-task reference baseline; the final functional verifier scored completed outputs but was not shown to the subject agent during repair. These results support archive-aware selection when evolution yields complementary specializations without consistent consolidation.
\end{abstract}

\input{sections/conference/01-introduction}
\input{sections/conference/02-background}
\input{sections/conference/03-framework}
\input{sections/conference/04-experiments}
\input{sections/conference/05-discussion}
\input{sections/conference/06-conclusion}

\input{sections/references}
\clearpage
\appendix
\input{sections/conference/appendix-a}

\end{document}

%% file: sections/conference/01-introduction.tex
\section{Introduction}

Design-verification debugging requires reasoning across regression logs, testbench code, RTL, configuration, simulation infrastructure, and domain-specific execution paths. An agent must find diagnostic anchors, trace a symptom to the logic that produced it, distinguish the root cause from the assertion or checker that detected it, and commit to an actionable answer within a bounded context and runtime. These tasks combine generic agent failures, such as speculative search or unbounded browsing, with hardware-specific failures involving launch and restore behavior, RTL update semantics, randomized traffic, resource accounting, and cross-layer testbench state.

We use \emph{lower-resource domain} to describe the relative scarcity of public task examples, debugging traces, tool-use demonstrations, and domain documentation compared with mainstream natural-language and software-engineering workloads. It does not refer to compute resources or model size. Prior hardware-language work provides relevant context: ChipNeMo~\citep{liu2023chipnemo} found benefits from domain adaptation for chip-design applications, while Revisiting VerilogEval~\citep{pinckney2024verilogeval} notes both the limited public representation of Verilog and the scarcity of hardware benchmarks.

This paper asks whether automatic harness evolution can compensate for part of that gap without changing model weights. We optimize a versioned agent package around a fixed subject model: the external runtime harness, together with the scaffold configuration it loads. The optimizer observes execution results and proposes new candidates. The same 12 tasks were used during search and final repeated evaluation, so the primary results measure optimization-set performance. Even in this favorable setting, evolution did not reliably fit the task set: the best candidate reached 10/12 pass@5, the best Success\(\geq\)2/5 result was 6/12, and the largest number of correct attempts was 21/60.

The main result is mixed but operationally important. Evolved harnesses completed far more attempts and solved more tasks across repeated trials, yet their gains were complementary rather than cumulative. A later candidate that inherited earlier rules and added causal-commit and turn-management mechanisms improved pass@5 and completion but lost first-trial and Success\(\geq\)2/5 coverage. A large candidate archive covered many more tasks than its incumbent. In this tested lineage, specialized behaviors did not consolidate reliably: no single evolved harness preserved all observed benefits.

This work makes three contributions. First, it presents an archive-aware harness-evolution framework evaluated across three distinct agent harnesses, with candidate isolation, activation checks, structured failure labels, and separate search and confirmation stages. Second, it reports a repeated optimization-set study showing large completion gains and broader pass@5 coverage, but only modest gains in correct attempts and no reliable fit to all 12 tasks. Third, it audits candidate manifests and traces to document an outcome-level failure to consolidate specialized behaviors within one evaluated lineage: useful mechanisms emerge in complementary candidates, archive coverage exceeds incumbent coverage, and inherited mechanisms do not reliably preserve predecessor behavior.

%% file: sections/conference/02-background.tex
\section{Background and related work}

An agent harness is the external runtime control layer that turns a language model into an executable agent by running the model--tool loop, managing context and state, mediating tool calls, enforcing execution and stopping policies, and returning results. We distinguish it from the underlying model, task environment, evaluator, and agent scaffold (prompts, tool descriptions, skills, and output schemas). Each experimental \emph{harness candidate} packaged the runtime harness with the scaffold configuration it loaded because the optimizer could modify both.

Prompt and language-model-program optimization provide an early foundation. DSPy~\citep{khattab2023dspy} treats language-model applications as optimizable declarative programs, and GEPA~\citep{agrawal2025gepa} uses trajectories and reflection to evolve prompts and combine lessons from a Pareto frontier. Agent-structure search exposes a wider surface: Automated Design of Agentic Systems~\citep{hu2024adas} has a meta-agent write new agent programs from an archive, while AFlow~\citep{zhang2024aflow} searches over code-represented workflows. These systems motivate optimizing more than a single instruction string.

Recent work modifies the harness directly. Meta-Harness~\citep{lee2026metaharness} gives a proposer access to previous harness code, scores, and traces. Self-Harness~\citep{zhang2026selfharness} mines weaknesses, proposes minimal changes, and validates them with regression tests. Recursive Harness Self-Improvement~\citep{lee2026recursive}, EvolveNet~\citep{nie2026evolvenet}, and HarnessDev~\citep{wu2026harnessdev} study iterative refinement, distributed aggregation, or broader harness construction and evolution. Our work differs in evaluating mixed RTL, testbench, functional-model, and regression evidence for root-cause analysis.

Two papers are especially relevant to scientific interpretation. Rethinking the Evaluation of Harness Evolution for Agents~\citep{wang2026rethinking} reports that reasonable changes to prompts, tools, budget management, verification, and finalization often improve robustness without consistently solving the remaining hard tasks. HarnessDev~\citep{wu2026harnessdev} similarly finds that feedback-set gains shrink or regress on held-out tasks, alongside run-to-run noise, revisions that help one outcome while hurting another, executor dependence, and mechanisms that were added but unreachable. Stop Comparing LLM Agents Without Disclosing the Harness~\citep{zhang2026disclosingharness} further argues that the harness is a material experimental variable. These findings motivate our repeated evaluation, broader-task replay, task-level reporting, and activation audit.

Comprehensive Verilog Design Problems~\citep{pinckney2025cvdp} introduces the CVDP benchmark for RTL design and verification in non-agentic and agentic settings. The closest harness-evolution comparison is Agentic Hardware Design as Repository-Level Code Evolution~\citep{yu2026horizon}, which reports recursive improvement on public RTL benchmarks, including CVDP, using executable acceptance feedback. That result concerns generation and repair under deterministic checks. Our primary setting instead requires post-failure causal diagnosis across heterogeneous artifacts, followed by exact matching of a format-normalized root-cause function. We analyze CVDP as a separate cross-benchmark case study rather than merge the two regimes.

%% file: sections/conference/03-framework.tex
\section{Harness-evolution framework}

\subsection{System overview}

The system separates a \emph{subject model}, which performs a debugging task, from an \emph{optimizer}, which proposes changes to the subject harness. Every candidate is materialized in an isolated directory, activated through its own configuration, evaluated on benchmark tasks, and recorded with its manifest, lineage, outputs, traces, scores, and failure state. The outer loop returns structured evidence to the optimizer and retains all candidates in an archive. The proposer model varied by campaign: Claude Opus 4.8 proposed changes in the MiniMax M3 and large branching searches, while Claude Opus 5 proposed changes in the DeepSeek V4 Pro and Kimi K3 searches. The proposer is distinct from the fixed subject model within each campaign. Figure 1 shows the architecture; Algorithm 1 gives the corresponding search and confirmation procedure.

\begin{figure}[h]
  \centering
  \includegraphics[width=0.98\linewidth]{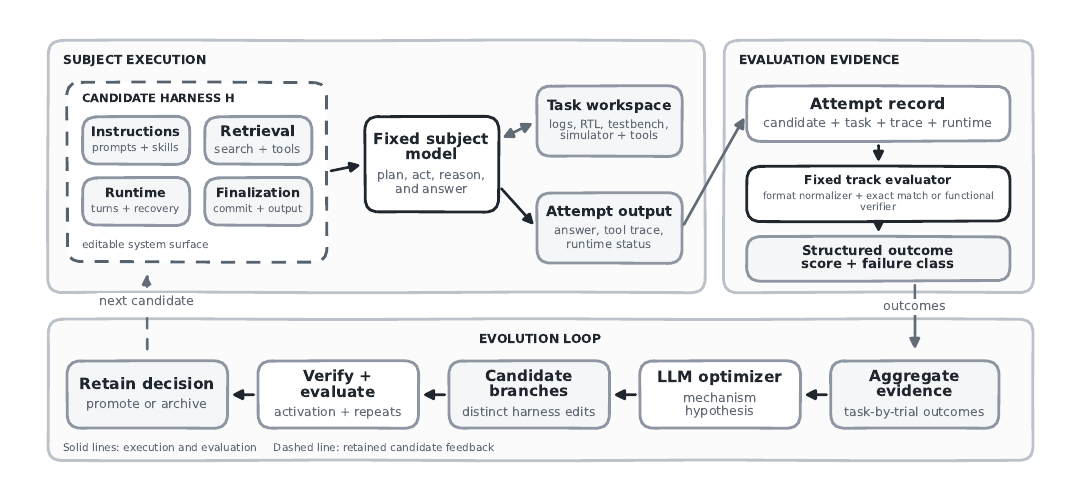}
  \caption{A candidate harness and task instantiate an isolated attempt around a fixed subject model. A separate outer loop proposes branches, verifies activation, evaluates candidates, confirms promising results, and retains both the incumbent and archive.}
  \label{fig:system-overview}
\end{figure}

\begin{paperalgorithm}{Harness evolution with archive-aware retention}
  \label{alg:harness-evolution}
  \par\noindent\hangindent=4.65em\hangafter=1\makebox[4.65em][l]{\textbf{Input:}}baseline harness \(h_0\); task set \(\mathcal{T}\); rounds \(K\); search trials \(R_s\); confirmation trials \(R_c\); promotion rule \(\operatorname{Promote}\).
  \par\noindent\hangindent=4.65em\hangafter=1\makebox[4.65em][l]{\textbf{Output:}}selected harness \(h\) and candidate archive \(\mathcal{A}\).
  \par\noindent\hangindent=1.95em\hangafter=1\hspace*{0.00em}\makebox[1.55em][r]{1.}\hspace{0.40em}\(h\leftarrow h_0\); \(\mathcal{A}\leftarrow\{h_0\}\)
  \par\noindent\hangindent=1.95em\hangafter=1\hspace*{0.00em}\makebox[1.55em][r]{2.}\hspace{0.40em}\(\mathrm{results}[h]\leftarrow\operatorname{Evaluate}(h,\mathcal{T},R_s)\)
  \par\noindent\hangindent=1.95em\hangafter=1\hspace*{0.00em}\makebox[1.55em][r]{3.}\hspace{0.40em}\textbf{For} \(k=1,\ldots,K\) \textbf{do}
  \par\noindent\hangindent=3.15em\hangafter=1\hspace*{1.20em}\makebox[1.55em][r]{4.}\hspace{0.40em}\(\mathrm{evidence}\leftarrow\operatorname{Summarize}(\mathrm{results},\mathcal{A})\)
  \par\noindent\hangindent=3.15em\hangafter=1\hspace*{1.20em}\makebox[1.55em][r]{5.}\hspace{0.40em}\(\mathcal{C}_k\leftarrow\operatorname{GenerateCandidates}(h,\mathrm{evidence})\)
  \par\noindent\hangindent=3.15em\hangafter=1\hspace*{1.20em}\makebox[1.55em][r]{6.}\hspace{0.40em}\textbf{For each} candidate \(c\in\mathcal{C}_k\) \textbf{do}
  \par\noindent\hangindent=4.35em\hangafter=1\hspace*{2.40em}\makebox[1.55em][r]{7.}\hspace{0.40em}\textbf{If} \(\operatorname{Active}(c)\), evaluate \(c\) for \(R_s\) trials; \textbf{else} record activation failure.
  \par\noindent\hangindent=3.15em\hangafter=1\hspace*{1.20em}\makebox[1.55em][r]{8.}\hspace{0.40em}\(\mathcal{P}_k\leftarrow\operatorname{SelectPromising}(\mathcal{C}_k,\mathrm{results})\)
  \par\noindent\hangindent=3.15em\hangafter=1\hspace*{1.20em}\makebox[1.55em][r]{9.}\hspace{0.40em}Confirm every \(c\in\mathcal{P}_k\) for \(R_c\) trials.
  \par\noindent\hangindent=3.15em\hangafter=1\hspace*{1.20em}\makebox[1.55em][r]{10.}\hspace{0.40em}\textbf{If} the best confirmed candidate \(c^\star\) satisfies \(\operatorname{Promote}(c^\star,h)\), set \(h\leftarrow c^\star\).
  \par\noindent\hangindent=3.15em\hangafter=1\hspace*{1.20em}\makebox[1.55em][r]{11.}\hspace{0.40em}\(\mathcal{A}\leftarrow\mathcal{A}\cup\mathcal{C}_k\) and store results, traces, activation status, and lineage.
  \par\noindent\hangindent=1.95em\hangafter=1\hspace*{0.00em}\makebox[1.55em][r]{12.}\hspace{0.40em}\textbf{Return} \(h,\mathcal{A}\).
\end{paperalgorithm}

We instantiated and evaluated the framework across three distinct agent harnesses: (1) an internal terminal coding-agent harness, (2) a custom Python harness built with the OpenAI Agents SDK~\citep{openai2026agentssdk}, and (3) a Codex CLI harness. Their editable boundaries differed: some runs primarily modified instructions and skills, while the custom implementation additionally exposed tool policy, subagent definitions, prompt construction, search and read wrappers, turn handling, runner behavior, and finalization. These configurations are not a matched comparison of agent frameworks; models, providers, and runtime conditions also differed. We interpret results within each experiment.

\subsection{Search, activation, and retention}

Prompt-constrained search targeted ungrounded queries, repetitive file reads, weak evidence tracking, late commitment, and answers lost to timeouts. Whole-harness search could also change executable search checks, read limits, turn management, runner behavior, scope controls, and finalization. Domain-context configurations supplied static documentation, retrieval, or increasingly hardware-specific proposer context. The domain information was given to the proposer or candidate harness, not added to benchmark tasks or the evaluator.

Branching generated sibling candidates from an incumbent or earlier branch and evaluated them independently. Because candidates often solved different tasks, the archive was a first-class output rather than a discard pile. It retained task-level outcomes, activation proofs, and lineage for later routing or synthesis. Promotion policies varied across exploratory runs, but the final study separates inexpensive discovery from five-trial confirmation.

Activation checks verified that a proposed rule or runtime change was present in the rendered subject harness before interpreting its score. This matters because a valid file on disk may never reach the model or entry point. Activation is necessary but not sufficient: a model can receive a rule and still fail to follow it. The evaluation therefore distinguishes malformed candidates, activation failures, provider and environment failures, subject timeouts, completed incorrect answers, and completed correct answers.

%% file: sections/conference/04-experiments.tex
\section{Experimental evaluation}

\subsection{Benchmarks, scoring, and protocol}

The primary benchmark contains 12 proprietary design-verification root-cause tasks, labeled T1--T12. Each task provides a materialized, read-only code workspace and asks the agent to identify the root-cause location and recommended change. The expected fixes span SystemVerilog, Verilog RTL, C++ functional models, and testbench code; failures include local guards and counters, RTL or interface-state updates, simulation and launch behavior, and cross-layer interactions. This is a within-domain stress test, not evidence of cross-domain generality.

The proprietary root-cause experiments use two scoring configurations. The exploratory DeepSeek V4 Pro and MiniMax M3 runs used a graded evaluator score \(s_t\) for each task \(t\) to measure overall answer quality, including improvements that a binary exact-match outcome would not capture. With the subject model fixed and only the harness changing within each run, score increases provided exploratory evidence of improved answer quality on selected evaluations; they do not isolate one harness component or establish that a gain persists on a broader task pool. A task counted as successful when \(s_t\geq0.8\). For an evaluation of \(K\) tasks, the DeepSeek comparisons report the mean graded task score, while Figure 4 reports both the number of successful tasks and the mean \(\bar{s}=K^{-1}\sum_{t=1}^{K}s_t\).

The repeated Kimi K3 study instead uses the stricter exact-match task-success criterion. Its evaluator checks the reference root-cause function. A short LM prompt only extracts and normalizes the predicted function identifier so that JSON, another structured representation, and ordinary prose share one scoring contract. The normalized identifier is then compared with the reference using exact match, making output format irrelevant to success. On two calibration sets, the selected normalizer returned the reference identifier for 45/46 and 34/34 outputs, respectively.

CVDP~\citep{pinckney2025cvdp} is analyzed as a separate cross-benchmark case study involving RTL generation, integration, or repair with executable compilation, simulation, and task-specific checks. Table 1 summarizes three evaluation settings across the two tracks. Their models, tasks, harnesses, evaluators, feedback, and reusable artifacts differ, so their scores are never combined or interpreted as matched effect sizes. The retained campaign used CVDP v1.1.0. The benchmark framework is distributed through NVlabs/cvdp\_benchmark~\citep{nvidia2026cvdpframework}, and the public dataset release is distributed as nvidia/cvdp-benchmark-dataset~\citep{nvidia2026cvdpdataset}.

\begin{table}[h]
  \centering
  \small
  \setlength{\tabcolsep}{3.5pt}
  \begin{tabularx}{\linewidth}{@{}>{\raggedright\arraybackslash}p{0.22\linewidth}>{\raggedright\arraybackslash}p{0.27\linewidth}>{\raggedright\arraybackslash}p{0.25\linewidth}>{\raggedright\arraybackslash}X@{}}
    \toprule
    \textbf{Evaluation track} & \textbf{Task type} & \textbf{Scoring criterion} & \textbf{Role in the study} \\
    \midrule
    Proprietary design-verification benchmark & 12 root-cause localization tasks & Format-normalized function exact match & Primary within-domain diagnostic evaluation \\
    Selected CVDP subset & 12 RTL generation and repair tasks & Executable functional verifier & Selected-subset phase of the cross-benchmark case study \\
    142-task CVDP follow-up & 142 RTL generation and repair tasks & Executable functional verifier & Baseline-to-evolved comparison in the cross-benchmark case study \\
    \bottomrule
  \end{tabularx}%
  \caption{Evaluation tracks and their roles in the study.}
\end{table}

An auxiliary experiment used eight search tasks and a four-task validation set excluded from search. Candidate selection used this validation set. We then replayed the selected candidate and baseline across all 12 tasks under matched settings. The four-task result is therefore a selection result rather than a held-out test, and the subsequent replay is not independent because it contains both the search and validation tasks.

\begin{figure}[h]
  \centering
  \includegraphics[width=0.98\linewidth]{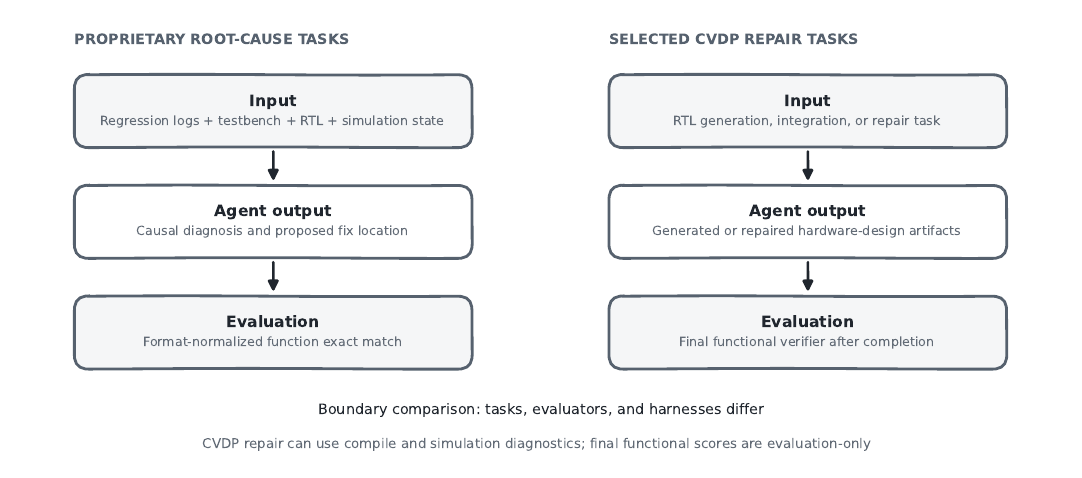}
  \caption{The primary benchmark evaluates causal diagnosis over mixed verification artifacts, while CVDP evaluates completed RTL outputs with executable functional checks. The final functional verifier was not shown to the subject agent during repair. The tracks support different claims and are not numerically interchangeable.}
  \label{fig:benchmark-contrast}
\end{figure}

We report four single-harness metrics: \emph{first-trial success} (tasks passed on designated trial zero), \emph{Success\(\geq\)2/5} (tasks passed in at least two of five trials), \emph{pass@5} (tasks passed at least once), and \emph{completion} (attempts reaching an evaluable result). First-trial success is not the conventional pass@1 estimator averaged over samples. We separately report \emph{archive union}, the tasks solved by any retained harness. Pass@5 and archive union measure breadth under resampling or portfolio selection; neither alone establishes reliable improvement in one harness.

\subsection{Primary results}

Table 2 reports the final repeated five-trial comparison around Kimi K3. All three rows use the same later, lower-concurrency rerun protocol; provider or runtime failures count as incomplete attempts. The evidence-grounded harness raised first-trial success from 2/12 to 6/12, pass@5 from 5/12 to 9/12, and completion from 34/60 to 58/60. Batched causal search completed all 60 attempts and reached 10/12 pass@5, but first-trial success and Success\(\geq\)2/5 were each 5/12, below the earlier candidate's 6/12. Total correct attempts increased more modestly, from 17/60 to 20/60 and 21/60. Thus later evolution improved some objectives while regressing others.

As a separate reference, a no-skills Codex + GPT-5.5 baseline completed 60/60 attempts and was reported at 5/12 pass@5. It used a different harness and an LM-judge-based scoring protocol, so it is not included in the matched Kimi K3 comparison in Table 2.

\begin{table}[h]
  \centering
  \small
  \setlength{\tabcolsep}{3.5pt}
  \resizebox{\linewidth}{!}{%
  \begin{tabular}{lccccc}
    \toprule
    \textbf{Harness} & \textbf{First trial} & \textbf{Success\(\geq\)2/5} & \textbf{Pass@5} & \textbf{Correct attempts} & \textbf{Completed} \\
    \midrule
    Kimi K3 baseline & 2/12 & 5/12 & 5/12 & 17/60 & 34/60 \\
    Automatically evolved evidence-grounded & \textbf{6/12} & \textbf{6/12} & 9/12 & 20/60 & 58/60 \\
    Automatically evolved batched-causal & 5/12 & 5/12 & \textbf{10/12} & \textbf{21/60} & \textbf{60/60} \\
    \bottomrule
  \end{tabular}%
  }
  \caption{Repeated evaluation results for the baseline and selected evolved harnesses.}
\end{table}

Figure 3 shows the task-level pass counts. Evidence-grounded search gained T3, T6, T7, T9, and T12 coverage but lost all five T10 trials and regressed on T2 and T11. Batched causal search recovered some T10 behavior and added T4, but lost T3 and was less reliable on T9. The two evolved harnesses together solved 11/12 tasks, while neither was uniformly strongest. T1 remained unsolved.

\begin{figure}[h]
  \centering
  \includegraphics[width=0.96\linewidth]{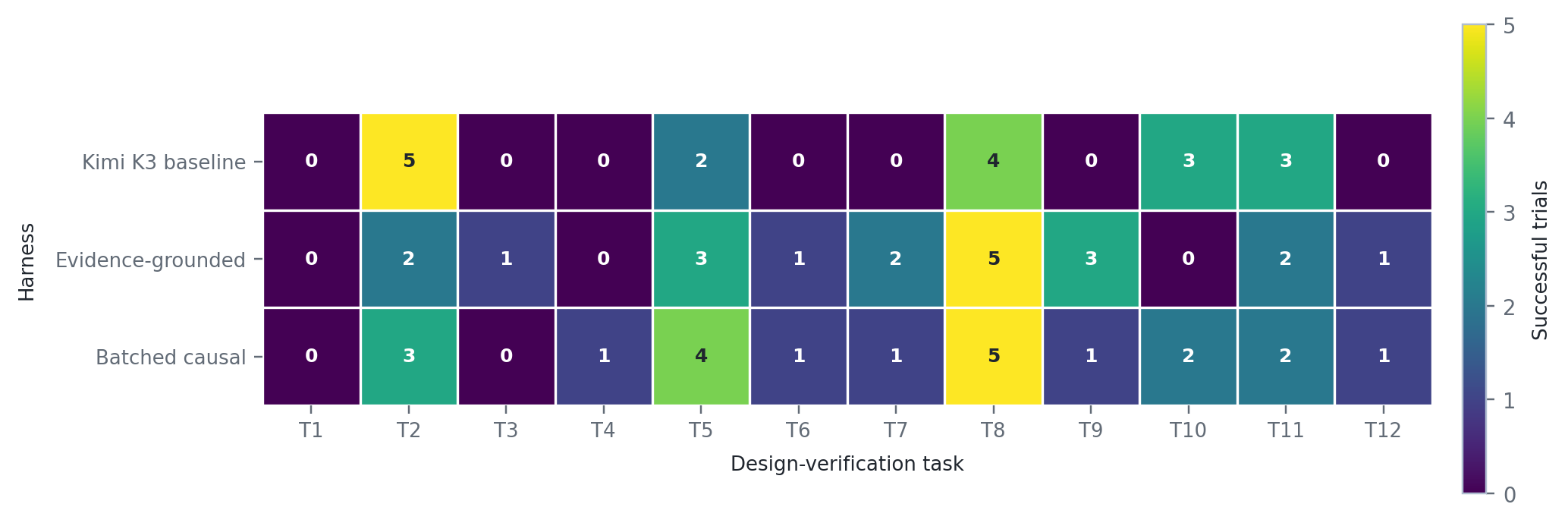}
  \caption{Successful trials out of five for each task and harness. Aggregate gains conceal regressions and complementary task-level strengths.}
  \label{fig:task-pass-matrix}
\end{figure}

Completion is itself a harness outcome: a diagnosis that never reaches an evaluable final answer has no operational value. The increase from 34/60 to 58/60 and then 60/60 shows that evolved runtime and finalization behavior was effective. The smaller change in correct attempts shows that completion and breadth account for more of the measured gain than repeated correctness on already completed attempts.

Selected performance peaks did not always persist across the broader pool. One DeepSeek harness rose from a mean task score of 0.375 to 0.725 on the four-task validation set excluded from search. In two all-12 replays, the candidate and baseline each solved 2/12 tasks; the candidate's mean task score was only 0.008 higher in one replay and 0.025 lower in the other. The selected validation gain therefore did not persist when replayed across the combined search and validation pool. Because that replay reuses both sets, it is not an independent test of generalization. A separate 20-replicate fixed-surface study likewise found no consistent advantage across task groups. These results support treating broader, repeated evaluations as primary evidence rather than relying on a selected peak.

An earlier MiniMax M3 experiment provides a longitudinal view of one search. The planned 300-generation run was deliberately stopped after 72 completed generations because it showed no further meaningful gains. With ten branches per generation and one trial per task, strict pass-set promotion changed the incumbent four times. The recorded incumbent rose from 1/12 task successes (\(\bar{s}=0.105\)) to 5/12 (\(\bar{s}=0.417\)). Completion initially fell from 9/12 to 2/12 at the first promotion, then recovered to 12/12. Figure 4 is a search trajectory rather than a learning curve: flat segments reuse the last accepted evaluation, and single-trial promotion does not quantify uncertainty.

\begin{figure}[h]
  \centering
  \includegraphics[width=0.94\linewidth]{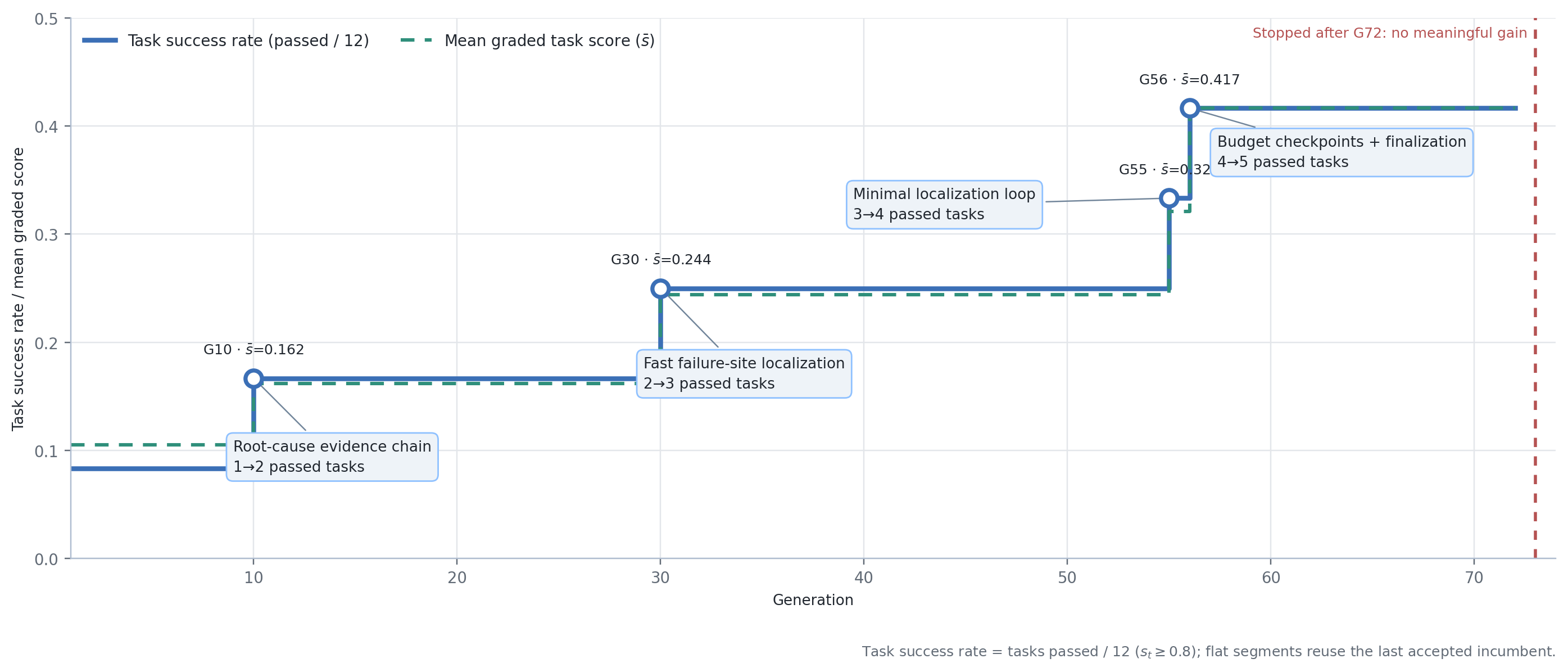}
  \caption{MiniMax task success rate and mean graded task score across 72 completed generations; the search was then stopped after no further meaningful gain. A task counts as successful when \(s_t\geq0.8\); \(\bar{s}\) averages the 12 graded task scores. The two final values both round to 0.417. Promotion labels identify the principal mechanism added at each accepted step.}
  \label{fig:minimax-trajectory}
\end{figure}

\subsection{What evolved, and what was executed}

The selected lineage concentrated on operational mechanisms rather than new hardware theories. Table 3 compresses the audited changes. The later candidates inherited earlier controls, so rows describe additions rather than independent treatments.

\begin{table}[h]
  \centering
  \small
  \setlength{\tabcolsep}{3.5pt}
  \begin{tabularx}{\linewidth}{@{}p{0.16\linewidth}p{0.42\linewidth}X@{}}
    \toprule
    \textbf{Stage} & \textbf{Principal addition} & \textbf{Intended effect} \\
    \midrule
    Evidence-first & Bounded orientation, literal search, anchored reads, forced finalization & Reduce speculative search and timeouts \\
    Coverage ledger & Observed-text and inspected-file ledger with search/read budgets & Balance premature fixation and uncontrolled exploration \\
    Evidence-grounded & Queries restricted to observed text; bounded sequential reads & Prevent invented queries and expensive paging \\
    Causal commit & Multiple evidence types; source-to-propagation-to-detection ordering & Avoid naming the checker as the cause \\
    Batched causal & Independent calls per turn plus standardized read windows & Gather more evidence within the turn budget \\
    \bottomrule
  \end{tabularx}%
  \caption{Audited mechanism lineage. Later stages inherit earlier controls.}
\end{table}

Activation proofs showed that the evidence-grounded search rules were present in a 28,788-character rendered prompt and that the later causal and batching rules were present in a 42,357-character prompt. The latter was approximately 47\% longer. This verifies delivery, not compliance or causality.

Two traces illustrate the distinction. On T9, evidence-grounded search passed 3/5 trials after a 0/5 baseline and one successful response localized a mixed-work-item synchronization error to the reference function. The active harness required a \texttt{LITERALS:} declaration before searching, but the preserved response omitted it. The attempt establishes correct localization despite incomplete rule compliance; it does not show that the grounding rule caused the success.

On T5, batched causal search passed 4/5 trials versus 2/5 for baseline. A successful response collected three witness types, rejected a downstream limit assertion as the detection site, committed to an upstream RTL launch-control transition, and emitted the required \texttt{COMMIT:} record. This directly demonstrates causal-commit compliance on that attempt, but it does not isolate batching or any single rule as the cause. Together, the cases show why both activation proofs and behavioral traces are required.

\subsection{Archive coverage and incomplete consolidation}

The strongest recurring pattern was a gap between discovering useful behaviors and combining them. A large branching experiment included an iteration-0 seed and generated ten branches in each of iterations 1--99, yielding 990 generated branches, of which 958 were evaluated. The baseline passed 1/12 tasks; the promoted incumbent reached 3/12 by iteration 8 and then remained unchanged through iteration 99. Only two candidates satisfied the strict promotion rule. The strongest unpromoted individual in the archive passed 4/12, while the archive union contained passes on 10/12 tasks. Additional search found specializations without consolidating them.

A second experiment adapted an existing proposer to the Codex subject harness. Across 65 iterations, its best-observed mean rose from approximately 0.26 to 0.29 near iteration 26 and then remained flat. Figure 5 contrasts task-level score activity in that run with free-form whole-harness evolution. Each cell is the best branch-attempt score for one task in an iteration, not the result of one harness. The free-form archive activated pass-level behavior on more rows, but those outcomes remained distributed across branches.

\begin{figure}[h]
  \centering
  \includegraphics[width=0.96\linewidth]{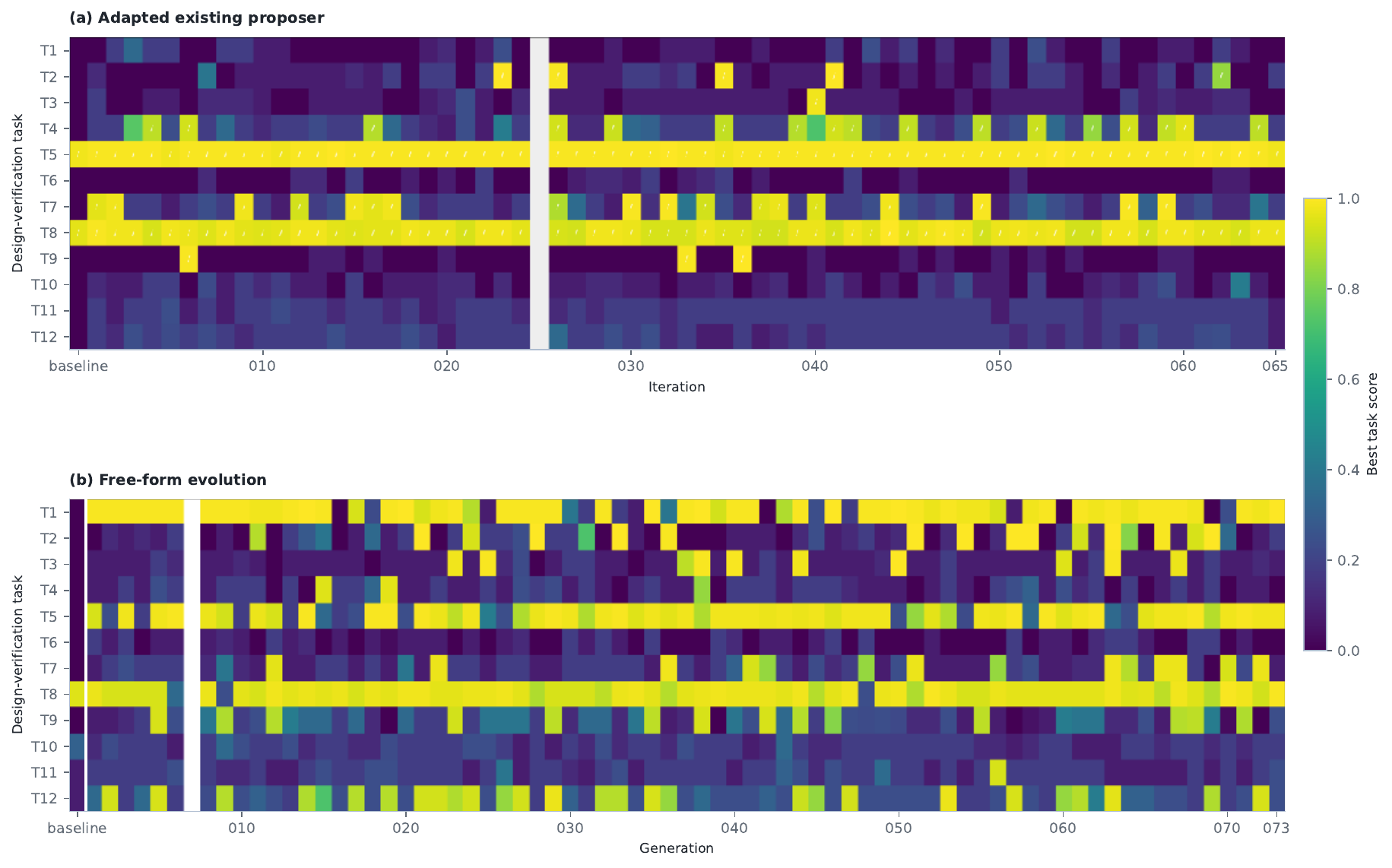}
  \caption{Best branch-attempt score by task and iteration for two evolution methods. The methods and run lengths differ, so this is a descriptive view of archive activity rather than a matched comparison.}
  \label{fig:method-activity}
\end{figure}

The repeated Kimi K3 lineage showed the same pattern under cleaner confirmation. Evidence-grounded search was strongest on first-trial success and Success\(\geq\)2/5, while batched causal search was strongest on pass@5 and completion. Their union reached 11/12 tasks. Figure 6 summarizes both archive gaps.

\begin{figure}[h]
  \centering
  \includegraphics[width=0.94\linewidth]{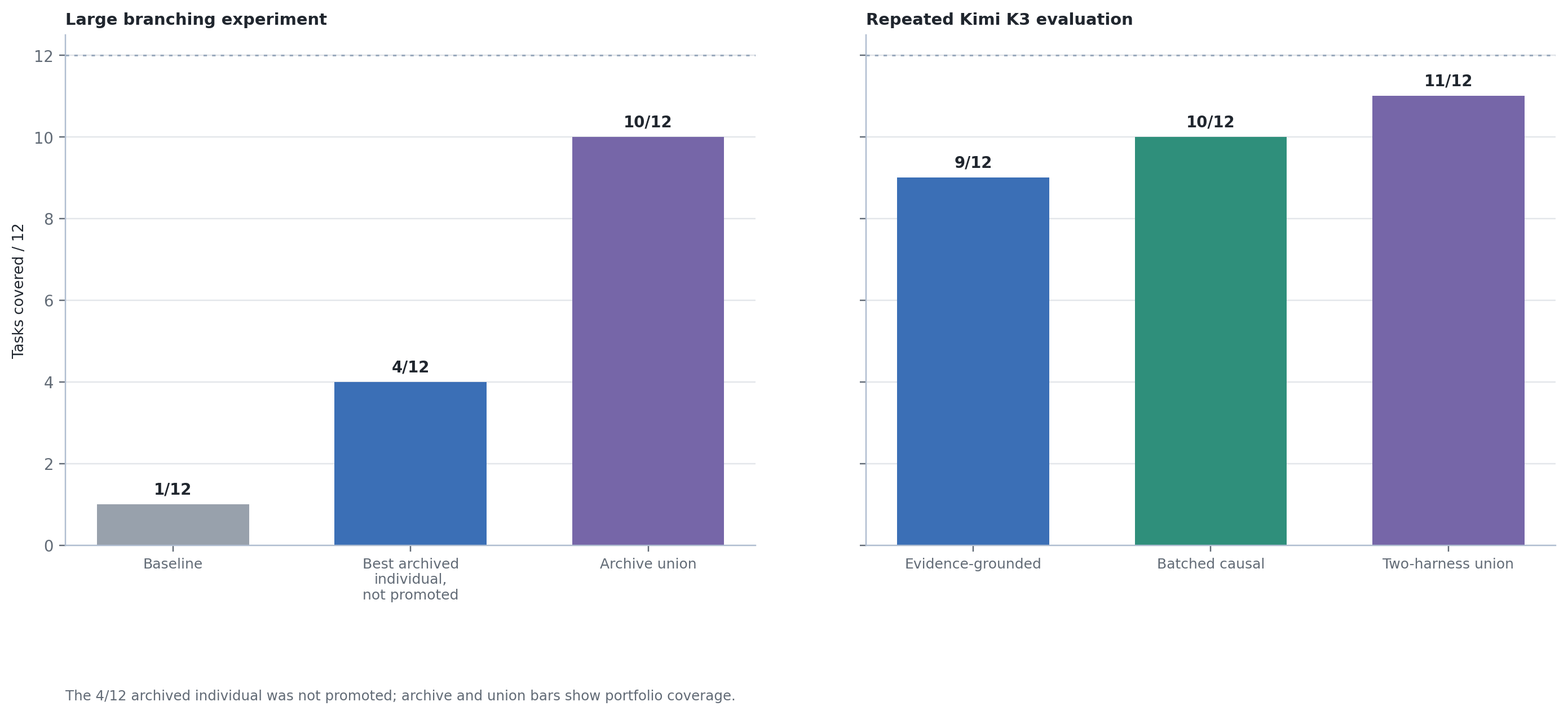}
  \caption{Candidate archives cover more tasks than individual harnesses. In the large branching experiment, the 4/12 bar is the best archived individual and was not promoted; the promoted incumbent remained at 3/12. Archive and union bars represent portfolio coverage, not the score of a composed generic harness.}
  \label{fig:candidate-compositionality}
\end{figure}

The later candidate explicitly inherited grounded-query and coverage mechanisms and added causal commitment, batched tool use, and turn management, but it did not preserve all measured advantages of its predecessor. We treat this as outcome-level evidence that the tested lineage failed to consolidate specialized behaviors into one consistently stronger harness. The artifacts do not identify whether the cause was optimizer choice, instruction interaction, subject-model compliance, or sampling variation.

Promotion is also a multi-objective statistical problem. A strict pass-set-superset rule can reject a useful candidate after a stochastic miss; an aggregate-score rule can promote easy new wins while losing a stable behavior. Repeated confirmation reduces but does not eliminate this ambiguity. A safer design retains complementary candidates, compares repeated task-level outcomes, permits explicit regression budgets, and considers routing among modular harnesses instead of continually extending one instruction set.

\subsection{Domain context and CVDP cross-benchmark case study}

\label{sec:cvdp-router}

We tested two distinct domain-context experiment families. In the first, static curated design-verification context or live retrieval from a specialized knowledge repository was available to the proposer. Each treatment peaked at 4/12 tasks, matching its standalone control. This result is separate from the empty-policy baseline in the word-density experiment below.

In the second family, an incremental meta-harness series increased domain-term density from about 11 to 25 terms per 1,000 words. The generated scaffolds changed vocabulary and referenced more domain artifacts, but the three treatment-level mean candidate pass rates remained between 10.3\% and 10.6\%, below the 14.6\% mean across four empty-policy baseline configurations. None of 58 evolved scaffolds exceeded the 3/12 best score also reached by an empty-policy baseline. Evolved scaffolds used about 30\% more input tokens and 28\% more agent time. Both families used one trial per task, and subject-side retrieval records for the live-retrieval experiment are incomplete. The supported claim is \emph{no demonstrated lift from the tested treatments}, not that domain knowledge is ineffective.

\begin{table}[h]
  \centering
  \small
  \setlength{\tabcolsep}{3.5pt}
  \begin{tabularx}{\linewidth}{@{}p{0.28\linewidth}p{0.18\linewidth}p{0.13\linewidth}X@{}}
    \toprule
    \textbf{Domain-context treatment} & \textbf{Mean candidate pass rate} & \textbf{Best result} & \textbf{Interpretation} \\
    \midrule
    Static curated context & 16.7\% & 4/12 & Matched standalone control \\
    Live knowledge retrieval & 17.5\% & 4/12 & Matched standalone control \\
    Low-density injection & 10.4\% & 3/12 & No reliable improvement \\
    Specificity instruction & 10.3\% & 3/12 & No demonstrated improvement \\
    Heavy domain context & 10.6\% & 2/12 & Content changed; score stayed flat \\
    \bottomrule
  \end{tabularx}%
  \caption{Domain-context treatments and best observed benchmark results.}
\end{table}

The CVDP cross-benchmark case study uses GLM-5.2 as the fixed subject model. On a selected subset, a lean baseline passed 4/12 functional verifiers, the best automatically evolved generic candidate passed 6/12, and the evolved generic archive covered 8/12. A later automatically evolved task-specific router using previously scored artifacts and task-family memories reached 10/12. A separate direct-clean baseline passed 2/12 under a different protocol, so 2/12 to 10/12 is not a controlled effect. In a 142-task follow-up, the automatically evolved defined-width repair harness passed 61 tasks, compared with 45 for the reference verification-loop baseline. During generation and repair, the subject agent could use ordinary compile, simulation, and testbench diagnostics and the harness's auxiliary review signal. The final benchmark functional verifier was run only after the candidate output was complete; its pass/fail result was not fed back to the subject agent for that task. The outer evolution process could use completed evaluations to select future harness candidates.

\begin{figure}[h]
  \centering
  \includegraphics[width=0.94\linewidth]{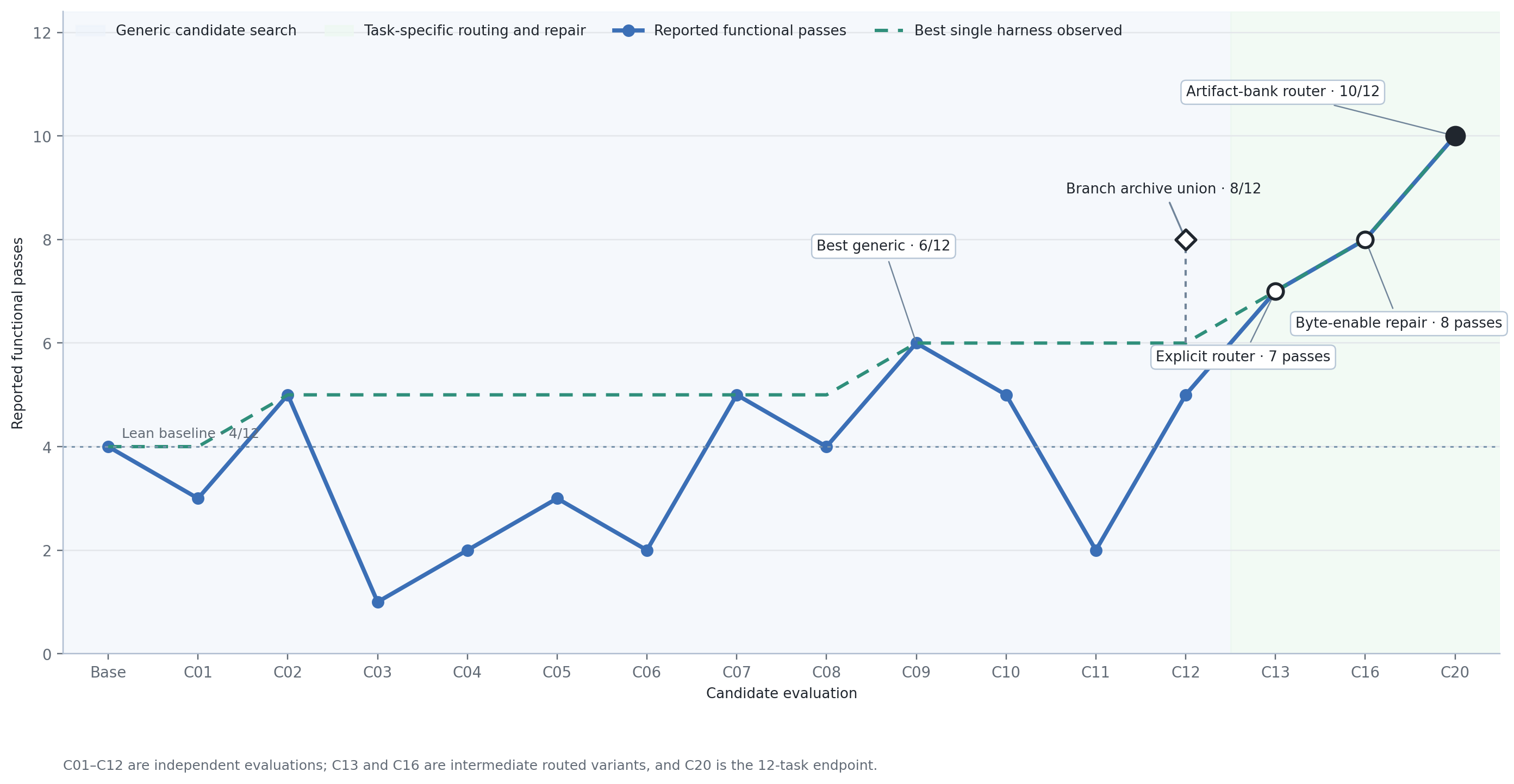}
  \caption{Functional-pass progression on the selected 12-task CVDP subset. The candidate harnesses were automatically evolved; later candidates use explicit routing and previously scored task artifacts, so the 10/12 endpoint is not a generic-harness result. Functional passes were measured after candidate outputs were complete.}
  \label{fig:cvdp-subset}
\end{figure}

\begin{table}[h]
  \centering
  \small
  \setlength{\tabcolsep}{3.5pt}
  \begin{tabular}{lcc}
    \toprule
    \textbf{CVDP harness condition} & \textbf{Subject LLM} & \textbf{Functional passes} \\
    \midrule
    Reference baseline: verification loop & GLM-5.2 & 45/142 (31.7\%) \\
    Automatically evolved: defined-width repair & GLM-5.2 & \textbf{61/142 (43.0\%)} \\
    \bottomrule
  \end{tabular}%
  \caption{Functional-verifier results for the reference baseline and automatically evolved repair harness on the 142-task CVDP follow-up.}
\end{table}

Table 5 compares the automatically evolved defined-width repair harness with the 142-task reference baseline recorded in the run artifacts. The evolved harness passed 16 additional tasks, a 35.6\% increase in functional passes over the baseline. Together with the selected-subset trajectory, the case study shows stronger outcomes with generation-time diagnostics, auxiliary review, and reusable task artifacts; it does not compare access to the final functional verifier during repair. Because the subject model, tasks, harness, evaluator, feedback, and artifacts differ from the proprietary study, it is neither a matched control nor an independent test of the same Kimi K3 harness.

%% file: sections/conference/05-discussion.tex
\section{Discussion and limitations}

\subsection{Interpretation and design implications}

The repeated results show improved operational robustness and task-specific outcomes, but not a coherent expansion of diagnostic capability. Evolved candidates completed more attempts and solved more tasks, while their task-level regressions, mechanism interactions, and coupled finalization changes prevent a stronger causal interpretation. This distinction is consistent with the external evidence from Rethinking the Evaluation of Harness Evolution for Agents~\citep{wang2026rethinking} and HarnessDev~\citep{wu2026harnessdev}: technically sensible edits can improve execution and robustness without moving every problem-solving outcome in the same direction.

The primary and CVDP tracks clarify where this limitation appears. In CVDP, compile, simulation, testbench, and auxiliary review signals available during generation can support localized repairs, while reusable task-specific artifacts can guide routing. The final functional verifier measured completed outputs; it was not a repair-time signal to the subject agent. Bespoke root-cause analysis in the lower-resource setting instead requires the model to decide which evidence matters, connect artifacts across the stack, and express a causal diagnosis before receiving an exact-match outcome. The experiments are not matched, but together they suggest that feedback structure and reusable experience condition how effectively evolution can exploit a fixed model.

The archive may therefore be more valuable than the final incumbent. Systems should preserve candidate manifests, activation proofs, task-level scores, trace summaries, and lineage; select or route among complementary harnesses; and keep archive-union coverage separate from single-harness performance. Discovery can use cheap trials and broad branching, but promotion should use repeated clean trials, task-level comparisons, and explicit regression budgets.

Critical constraints should be executable when possible. Search-evidence checks, read limits, activation verification, and forced finalization can be instrumented rather than left entirely to instruction following. This reduces the model's compositional burden and makes violations observable. Evaluation should also keep provider outages, inaccessible workspaces, malformed candidates, inactive policies, subject timeouts, and incorrect completed diagnoses in separate categories. Collapsing them into one failure signal encourages the optimizer to learn from noise.

\subsection{Limitations}

The strongest repeated root-cause result covers only 12 proprietary optimization tasks and is too small for broad claims about design verification. We did not scale that benchmark because no evolved harness reliably fit even the existing set: the best candidate reached 10/12 pass@5, the best Success\(\geq\)2/5 result was 6/12, and the largest number of correct attempts was 21/60. Scaling before resolving the selection and consolidation failure observed in this lineage would have increased cost without addressing the observed failure. The auxiliary validation-set gain did not persist in the full 12-task replay, but that replay contains both search and validation tasks and is not an independent test. The CVDP cross-benchmark case study reports improvements within its selected-subset sequence and over a 142-task reference baseline; it does not measure generalization of the same Kimi K3 harness.

The selected harnesses differ in multiple instructions and runtime controls, so their score changes cannot be assigned to one edit. Completion and diagnostic behavior changed together. Five trials per task reduce but do not remove sampling variation, and earlier exploratory runs used single-trial promotion. The normalizer calibration found one mismatch across 80 audited outputs.

The word-density domain-context study used one trial per task, four empty-policy baselines, and coarse 1/12 increments; the separate live-retrieval study retained incomplete subject-side retrieval records. The CVDP subset was selected, its 10/12 endpoint used task-specific routing, and its direct-clean baseline followed a different protocol. Provider and environment failures affected earlier exploratory reruns; all three primary Kimi K3 rows use the same later, lower-concurrency rerun protocol, with provider or runtime failures counted as incomplete. Finally, the lower-resource characterization is based on public-data and benchmark scarcity.

%% file: sections/conference/06-conclusion.tex
\section{Conclusion}

Harness evolution improved how a fixed model used its available capabilities on design-verification debugging. Evidence-grounded and batched causal harnesses increased completion, first-trial success, and pass@5 over the baseline, and their union solved 11/12 tasks. The optimizer proposed grounded search, bounded reads, evidence tracking, causal commitment, turn management, and finalization controls that were present in active harnesses and sometimes visible in successful traces.

The central empirical finding was a failure to consolidate specialized behaviors within the evaluated lineage. Useful changes produced complementary task profiles, the archive covered more tasks than any incumbent, and a larger combined harness did not dominate its predecessor. More domain terminology also changed scaffolds without demonstrating better outcomes. The separate GLM-5.2 CVDP cross-benchmark case study showed stronger functional outcomes with generation-time diagnostics, auxiliary review, and reusable task artifacts, including a 35.6\% increase in functional passes for the automatically evolved defined-width repair harness over the 142-task reference verification-loop baseline. The subject agent did not receive the final functional-verifier result during repair.

Automatic harness evolution is therefore promising as a method for discovering operational safeguards and specialized candidates in lower-resource hardware domains, but it is not yet a substitute for domain competence or deliberate system design. Near-term systems should treat the archive as an output, verify mechanism activation and execution separately, confirm promotions with repeated task-level evaluation, and reduce the burden on the LLM by enforcing critical constraints in code.

%% file: sections/conference/appendix-a.tex
\section{Supporting experiments and audit detail}

\subsection{Additional framework views}

\begin{figure}[h]
  \centering
  \includegraphics[width=0.98\linewidth]{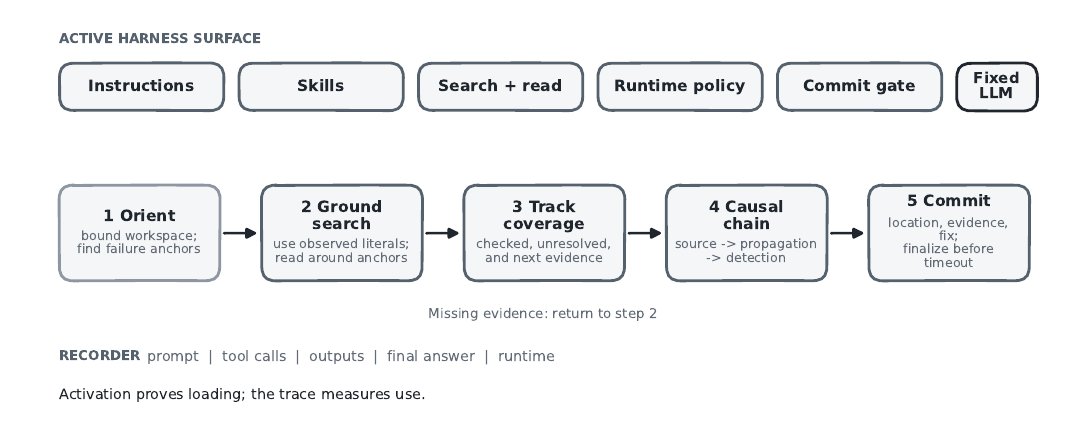}
  \caption{Evidence acquisition, coverage tracking, causal analysis, and final commitment within one subject attempt.}
  \label{fig:subject-loop}
\end{figure}

\begin{figure}[h]
  \centering
  \includegraphics[width=0.98\linewidth]{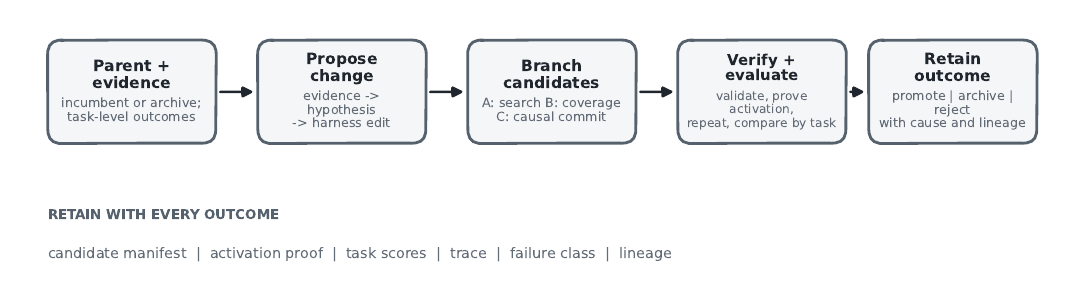}
  \caption{Candidate branches pass through materialization, static validation, activation checks, evaluation, and experiment-specific retention.}
  \label{fig:candidate-lifecycle}
\end{figure}

\subsection{Benchmark detail and outcome labels}

\begin{table}[h]
  \centering
  \small
  \setlength{\tabcolsep}{3.5pt}
  \begin{tabularx}{\linewidth}{@{}p{0.28\linewidth}p{0.16\linewidth}X@{}}
    \toprule
    \textbf{Design-verification task subfamily} & \textbf{Tasks} & \textbf{Typical failure or reasoning pattern} \\
    \midrule
    Local testbench or resource accounting & T1, T2, T3, T4, T5, T7 & Guards, counters, resource accounting, and tracing a local symptom to its producer \\
    RTL or interface-state update & T8, T12 & Assignment semantics, state transitions, and calling-logic behavior \\
    Simulation, launch, or cross-layer behavior & T6, T9, T10, T11 & Stimulus, checking, memory modes, restore behavior, and multi-layer failures \\
    \bottomrule
  \end{tabularx}%
  \caption{Provisional design-verification task subfamilies represented in the benchmark.}
\end{table}

The grouping is retrospective and was not shown to agents. Subcategories overlap; it is used for descriptive analysis rather than as a benchmark label.

\subsection{Domain-context and cross-model detail}

\begin{table}[h]
  \centering
  \small
  \setlength{\tabcolsep}{3.5pt}
  \resizebox{\linewidth}{!}{%
  \begin{tabular}{lccc}
    \toprule
    \textbf{Treatment} & \textbf{Knowledge treatment} & \textbf{Mean candidate pass rate} & \textbf{Best result} \\
    \midrule
    Static curated DV context & Documentation in proposer context & 16.7\% & 4/12 \\
    Live DV retrieval & Proposer could retrieve specialized knowledge & 17.5\% & 4/12 \\
    Low-density injection & About 11 terms per 1,000 words & 10.4\% & 3/12 \\
    Specificity instruction & About 18 terms per 1,000 words & 10.3\% & 3/12 \\
    Heavy DV context & About 25 terms per 1,000 words & 10.6\% & 2/12 \\
    \bottomrule
  \end{tabular}%
  }
  \caption{Domain-context treatments and their best observed scores.}
\end{table}

\begin{table}[h]
  \centering
  \small
  \setlength{\tabcolsep}{3.5pt}
  \resizebox{\linewidth}{!}{%
  \begin{tabular}{lccccc}
    \toprule
    \textbf{Configuration} & \textbf{Evaluated configurations} & \textbf{Mean pass rate} & \textbf{Best score} & \textbf{Input tokens} & \textbf{Agent time} \\
    \midrule
    Empty-policy baselines & 4 & 14.6\% & 3/12 & 5.3--6.7M & about 1,250 s \\
    Evolved domain scaffolds & 58 & n.r. & 3/12 & 7.4--8.2M & about 1,600 s \\
    \bottomrule
  \end{tabular}%
  }
  \caption{Empty-policy baselines in the incremental domain-context experiment.}
\end{table}

The domain-context values were reconstructed from experiment summaries rather than raw per-attempt artifacts and are treated as descriptive. The pooled mean for the 58 evolved scaffolds is not reported because the per-treatment candidate counts and unrounded scores needed to reconstruct it were not retained; n.r. denotes this unreconstructed aggregate, not a zero score. The three treatment-level means are shown in the preceding table.

\begin{table}[H]
  \centering
  \small
  \setlength{\tabcolsep}{3.5pt}
  \begin{tabular}{lcc}
    \toprule
    \textbf{Model} & \textbf{Passed attempts / 36} & \textbf{Timeouts} \\
    \midrule
    GLM 5.2 & 12/36 & 6 \\
    Kimi K2.6 & 6/36 & 17 \\
    Nemotron Ultra & 4/36 & 4 \\
    MiniMax & 2/36 & 23 \\
    \bottomrule
  \end{tabular}%
  \caption{Cross-model descriptive baseline results.}
\end{table}

The cross-model runs used an older configuration and are not controlled comparisons with the final Kimi K3 evolution results.

\subsection{CVDP inventory}

The full follow-up evaluated 142 selected tasks from CVDP v1.1.0: 49 \texttt{cvdp\_agentic} and 93 \texttt{cvdp\_copilot} tasks. The preserved dataset directory contains 151 task directories, so directory enumeration would overstate the evaluated set. The 12-task subset corresponds to ordinals 1--12 of the selected tasks.

Figure 5 is generated from archived heatmap rasters and is descriptive because raw numeric matrices were not retained.